# Future state prediction based on observer for missile system

W. K. Smithson, X]b\i U Wang
Aerospace Engineering,
University of Nottingham, UK
9a U]`.`k Ub[ I ]b\i U$( 4 [ a U]¨Wta

**ABSTRACT**

Guided missile accuracy and precision is negatively impacted by seeker delay, more specifically by the delay introduced by a mechanical seeker gimbal and the computational time taken to process the raw data. To meet the demands and expectations of modern missiles systems, the impact of this hardware limitation must be reduced.

This paper presents a new observer design that predicts the future state of a seeker signal, augmenting the guidance system to mitigate the effects of this delay. The design is based on a novel two-step differentiator, which produces the estimated future time derivatives of the signal. The input signal can be nonlinear and provides for simple integration into existing systems.

A bespoke numerical guided missile simulation is used to demonstrate the performance of the observer within a missile guidance system. Both non-manoeuvring and randomly manoeuvring target engagement scenarios are considered.

## NOMENCLENTURE

| | | |
|---|---|---|
| $x$ | State vector | |
| $y$ | Output vector | |
| $u$ | Input vector | |
| $\dot{\lambda}$ | Line of sight rate | [rad·s$^{-1}$] |
| $\hat{\dot{\lambda}}$ | Predicted line of sight rate | [rad·s$^{-1}$] |
| $v(t)$ | Observer input signal | |
| $x_{i,j}$ | Observer state variables | |
| $k_i$ | Observer tuning variables | |
| $\varepsilon$ | Observer perturbation function | |
| $\Delta$ | Observer time offset | [s] |
| $d_{miss}$ | Miss distance | [m] |
| $r_t$ | Target position vector | [m] |
| $r_m$ | Missile position vector | [m] |

## 1. INTRODUCTION

A missile's guidance navigation and control (GNC) subsystem is concerned with determining the trajectory and the force input required to fulfil a mission's guidance commands. The GNC concept was first introduced in WWII as the Germans developed the V1 and V2 guided missiles.[1]

In the modern era there is an increasing demand for accuracy, better performance against improving counter measures and better performance against stealth capabilities. With this increasing demand on missile systems, more advanced guidance laws are being required.

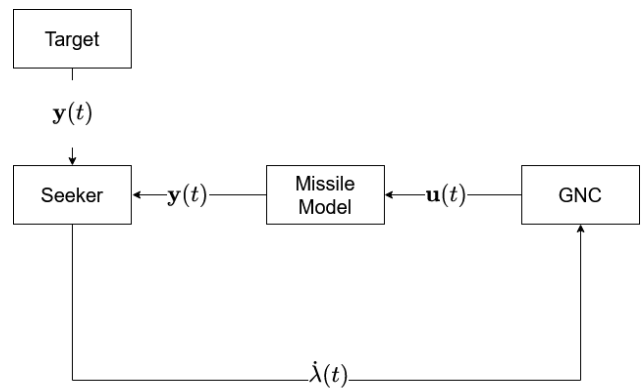

**Figure 1: Guidance navigation and control loop**

A missiles GNC suite consists of several components as illustrated in Figure 1. The seeker is a sensor that provides the guidance algorithm information relating to the target's positions. This normally is in the form of relative angular position of the target but can also include other information such as range.[2] The guidance system implements this guidance law and produces the trajectory requirements which are then interpreted by the auto pilot system. This system will produce control surface deflections to achieve the desired trajectory from the missile.

The guidance law implemented in almost all the world's guided missile systems is known as proportional navigation.[1]

The paper, *The Fundamentals of Proportional Navigation*[3], provides a brief overview of this technique in the context of a satellite interceptor system.

In summary, the concept is that if two objects with different trajectories maintain the same angle of bearing between their respective trajectories, the objects will eventually collide.

Although this paper presents a good introduction to the technique, the application to a satellite interceptor is an idealistic scenario. It does not account for dynamic manoeuvring of a normal airborne target. More complex guidance laws exist that can account for this and offer better performance.[4] These guidance laws generally offer superb theoretical performance for accuracy and efficiency but in practical application, their performance is often degraded by implementation constraints such as hardware limitations.

A prominent issue introduced in practical applications is the lag in the line of sight (LOS) measurement. This is introduced by the seeker gimbal and processing of the signals it produces. With the requirement to intercept supersonic targets, this delay contributes to a significant increase in miss distance[5]; minimisation of this effect is explored in a number of papers using a range of techniques.\cite{SeekDelay,7813436}[6, 7]

In modern control theory, dynamical systems are often represented as a system of first order differential equations defined by a set of variables known as the state vector. These equations combined with input and output behaviours of the system provide a comprehensive system model. This representation is commonly known as a *state-space model* in control theory.

$$\dot{x} = Ax + Bu$$
$$\dot{y} = Cy + Du \qquad (1)$$

It is often assumed in control theory that the state vector, $x$, is known but this is often not the case in real systems. When this occurs, it becomes necessary to design a system to estimate the state vector. Estimation of the state variable is called *observation* and the implementation of such a system is referred to as a *state observer*. Commonly, this is implemented as a computer program due to the current pervasiveness of microchips and the ease of use these devices offer compared to other solutions such as analogue electronics.

A full order state observer provides an estimation of all states in a system. This is commonly implemented by producing a mathematical model that emulates the output of a real system. The error between the true system output and the model is then used within a feedback loop to correct the model.

If the observer does not observe all variables contained in the state vector, it is referred to as a reduced order observer.[8] This type of system is the basis of this dUdYf.

This paper presents a novel solution to the seeker delay problem based on a predictor observer originally proposed by Wang and Lin[9].

The new observer offers compelling features for applications in reducing the seeker delay.

It takes an input signal and produces estimated future time derivatives of the signal. The input signal can be unknown and nonlinear. Furthermore, it rejects high frequency noise introduced by measurement sensors in real applications. These features suggest it should be suitable for use within a proportional guidance simulation.

Previous solutions to the seeker delay problem achieved effective results, however they rely on relatively more complex methods[7]. A solution based on the design set out in this dUdYf would offer much more simplicity. Building on the findings it might be possible to improve the absolute performance of seeker delay correction algorithms, further reducing miss distance within missile systems that employ seekers.

## 2. METHODOLOGY AND MAIN RESULT

To study the performance of the observer a bespoke numerical guided missile simulation employing a proportional navigation guidance algorithm replicating a ground to air engagement was used. First, a baseline result was obtained from the simulation without the use of the observer. The guidance algorithm in the model was then then augmented with the observer to remove the seeker delay.

A range of statistical and practical metrics have been used to compare these results to determine if the algorithm can:
1. Accurately track the future dynamics of a missile system when engaging a target.
2. Determine whether this correlates to an improved performance when the observer is used to remove delays in LOS measurements provided to the guidance algorithm.

This process was first completed using a non-manoeuvring target. This was important to effectively calibrate the observer for the dynamics of the simulation and provide an overview of the system without having to compensate for an overly complex target model. A second set of results was obtained using a randomly manoeuvring target to assess the performance against a more realistic target. A Monte-Carlo method was used to reduce the effects of bias introduced through the target selection when the randomly manoeuvring target was used.

## 2.1 Observer design

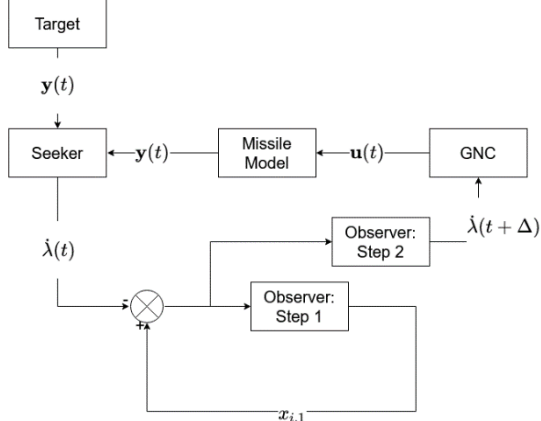

**Figure 2: GNC loop augmented with observer**

The observer used in this pUdYf is based on a two-step differentiator for delayed signal. Figure 2 shows how it was integrated in the standard guidance loop presented in Figure 1 in a block diagram form.

$$\dot{x}_{1,1} = x_{2,1} + \frac{k_1}{\varepsilon}(v(t) - x_{1,1})$$
$$\dot{x}_{2,1} = x_{3,1} + \frac{k_2}{\varepsilon^2}(v(t) - x_{1,1})$$
$$\dot{x}_{3,1} = x_{4,1} + \frac{k_3}{\varepsilon^3}(v(t) - x_{1,1})$$
$$\dot{x}_{4,1} = x_{3,1} + \frac{k_4}{\varepsilon^4}(v(t) - x_{1,1})$$

$$\dot{x}_{1,2} = x_{2,2} + \left(\frac{1}{6}\frac{k_4}{\varepsilon^4}\Delta^3 + \frac{1}{2}\frac{k_3}{\varepsilon^3}\Delta^2 + \frac{k_2}{\varepsilon^2}\Delta + \frac{k_1}{\varepsilon^1}\right) \times (v(t) - x_{1,1})$$
$$\dot{x}_{2,2} = x_{3,2} + \left(\frac{1}{2}\frac{k_4}{\varepsilon^4}\Delta^2 + \frac{k_3}{\varepsilon^3}\Delta + \frac{k_2}{\varepsilon^2}\right)(v(t) - x_{1,1})$$

$$\dot{x}_{3,2} = x_{4,2} + \left(\frac{k_4}{\varepsilon^4}\Delta + \frac{k_3}{\varepsilon^3}\right)(v(t) - x_{1,1})$$
$$\dot{x}_{4,2} = \frac{k_4}{\varepsilon^4}(v(t) - x_{1,1}) \qquad (2)$$

The observer consists of two systems of first order ordinary differential equations (ODE) that describe the two steps of the observer used to estimate the future state of the input signal, $v(t)$. The first step state variables, $\dot{x}_{i,1}$, estimates the current derivatives of the input signal. The second step uses the values obtained from step one to produce the second step state, $\dot{x}_{i,2}$. These state variables estimate the input signal and its derivatives at $t + \Delta$. Thus:

$$\begin{aligned} x_{i,2} &\to v^{(i)}(t + \Delta) \\ i &= \{1, 2, 3, 4\} \end{aligned} \qquad (3)$$

In this application the input signal $v(t)$, is the current LOS rate. The observer will predict the future derivatives of this signal, which is stored in the state variables $x_{n,2}$.

The variables $k_n$ are the tuning variable and effect the stability of the observer. These are selected to optimize the performance of the observer. For this application, a one-factor-at-a-time method will be used to select these parameters until a suitable stability and convergence time, with respect to the true seeker output, has been achieved. This will be determined heuristically, thus the results from this pUdYf do not represent the optimal performance of this observer.

An issue that occurs with design of this observer is a peaking phenomenon that causes significant overshoot at the start of the observation period. This is overcome by implementing a perturbation parameter $\varepsilon$, that controls the gain of the observer and is designed limit the peaking phenomenon. Finally, $\Delta$ is used to determine how far into the future the observer will predict. Over the course of a simulation this variable is held constant, but it can be adjusted to account for varying seeker delays.

## 2.2 Numerical simulation

The numerical simulation used for this pUdYf was a bespoke solution that was designed to emulate the dynamics of a guided missiles as accurately as possible. It also provides a suitable interface to integrate the new observer as well as tools to analyse the simulations performance. To ensure that it produced valid results. Development followed standard practices for flight and missile simulations as well as using aerodynamic and thrust data from *MIL-HDBK-1211*. [10, 11]

The simulation uses a nonlinear five degrees of freedom (5 D.O.F) model. Guided missile generally skid-to-turn rather than rely on bank angle to produce a lateral acceleration (as used in fixed-wing aircraft) so roll motion can be neglected in the simulation. This simplifies the equations of motion for the model without significantly reducing accuracy. It will also significantly simplify the implementation of guidance laws. Furthermore, it reduces the detail of the aerodynamic model because there is no requirement for the roll dynamics of the air frame to be modelled. Finally, it delivers practical benefits by reducing the execution time of the simulation program.

The aerodynamics were implemented using coefficient equations for the forces and moments. The coefficients themselves were obtained by interpolating from a discrete a simple aerodynamic database that approximates the coefficients as linear equations dependant on angle of attack (AOA) and represents a generic model. These vary by Mach number to provide accurate aerodynamic forces as the atmosphere and velocity change throughout the simulation.

The model was further simplified by assuming the airframe has a cruciform symmetry. This means that the pitch and yaw aerodynamics are identical, reducing the database required to produce an accurate aerodynamic model.

This method has limitations as stall characteristics are not modelled and it is not fully representative of a real airframe. It is acceptable for this application because the scenarios modelled do not require the missile to approach the stalling regions and emulating a real missile is not within the scope of this dUdYf; it will provide a realistic response for a generic airframe.

Thrust characteristics are also modelled using a data table. Atmospheric conditions are included in the simulation. This uses the standard atmosphere model to generate the appropriate parameters.

The seeker measures the line of sight rate using the exact state of the target and missile to provide an initial true LOS rate, $\dot{\sigma}$. A delay is then imposed on the signal by introducing a first order transfer function:

$$\dot{\lambda}(t + \tau) = \dot{\lambda}(t)e^{-\tau/\Delta} + \dot{\sigma}(t)\left(1 - e^{-\tau/\Delta}\right) \quad (4)$$

The resulting signal is the delayed LOS signal, $\dot{\lambda}$. The observer is used to predict the future state of this signal and provide the guidance algorithm an accurate position for the target.

## 2.3 Simulated scenarios

The simulation models two different ground to air engagement scenarios using two different targets.

The first target is a non-manoeuvring target that holds a constant altitude and travels at a constant velocity towards the missile launch site. In a realistic engagement the target will generally try and outmanoeuvre the missile the resulting motion is not known by the GNC suite beforehand. To simulate a more realistic engagement the second target is randomly manoeuvring.

This second target has the same dynamics as the non-manoeuvring target except it replaces the zero-vertical velocity with sinusoidal vertical velocity with a magnitude of 5m/s and frequency of 3 rad/s. The initial phase is randomly selected at the beginning of the simulation resulting in a random engagement each time the simulation is run. This will provide a suitable approximation to gauge the observer's performance against a realistic target. The combination of these two target models provides a suitable range of engagements to assess the observer.

## 2.3 Performance metrics

A range of techniques will be used to analyse the performance of the observer.

The root mean squared error (RMSE) will quantify the performance of the observer across the entire simulation.

$$RMSE = \sqrt{\sum_{i=1}^{n} \frac{\left(\hat{\dot{\lambda}}(t + \Delta) - \dot{\lambda}(t + \Delta)\right)^2}{n}} \quad (5)$$

This is a standard technique used to measure the aggregate residuals between values predicted by a model and the true value. This will provide a useful quantity for comparing the effect different parameters have on the observer across different simulation runs. This will enable the assessment of the observer based on different inputs to model and different observer parameter i.e. prediction time difference.

The previous techniques provide the ability to measure the performance of the observer from a theoretical sense, but they do not indicate the practical performance. Two common metrics used in missile design that provide an indication of how the observer will perform:

- Miss distance
- Commanded acceleration

Miss distance is defined as the smallest Euclidean distance between the target and the missile:

$$d_{miss} = min|\overline{r_t} - \overline{r_m}| \quad (6)$$

As the miss distance decreases the chance to successfully intercept the target increases. Therefore, this provides a useful measure on how successful a guidance algorithm will be in each engagement.

Commanded acceleration defines the magnitude of the acceleration requested by guidance algorithm and is represented by the magnitude of the control surface deflections. A small commanded acceleration not only indicates that the algorithm is efficient, but also reduces the performance advantage a missile must possess to intercept a target successfully. Even if an algorithm possesses outstanding miss distance characteristic, if will not be effective if a missile cannot produce the required performance.

### 2.4 Monte-Carlo simulation

To eliminate selection bias from the randomly manoeuvring target a Monte-Carlo simulation will be used. The simulation will run multiple times with a randomly generated targets and the average of these samples will be taken to study the effects of the observer for this type of engagement.

Multiprocessing techniques have been used to optimise the simulation execution time, but the simulation was limited to 25 samples for the eight delays explored, for a total of 200 simulations so that the results can be obtained in a reasonable time frame.

Miss distance will be sampled for a range of delays from 0.025 to 0.35 seconds. The mean for each time delay has been calculated and the standard deviation will also be used to judge the performance of the observer.

## 3. ANALYSIS AND DISCUSSION

### 3.1 Non-manoeuvring target

As demonstrated in Figure 3, it was possible to achieve a stable simulation using the observer to remove seeker delay. Based on the data presented in Table 1 the missile was able to intercept the target efficiently and the result of adding the observer was a 54% reduction in the miss distance of the missile compared to the same simulation without the correction.

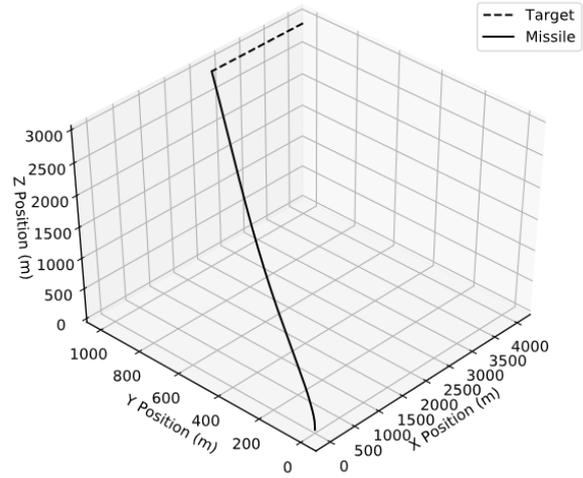

**Figure 3: Trajectories of the missile and target during the simulation**

**Table 1: Non-manoeuvring target performance statistics**

|  | LOS rate RMSE | Miss distance |
|---|---|---|
| Zero delay | 0 | .001m |
| Uncorrected | .0029 | .175m |
| Corrected | .0004 | 080m |
| % difference | -54 | -800 |

The observer also demonstrated an excellent ability to track the LOS rate outputted by the seeker with an 800% reduction of the RMSE - again compared to the uncorrected signal. This suggests that the predicted signal is much closer to the true LOS rate than the original delayed signal. This prediction accuracy is the main factor contributing to the large reduction in miss distance however it was not possible to achieve the same miss distance as was achieved with zero delay; likely due to the observer struggling with the rapidly changing los rate in the final milliseconds of the engagement.

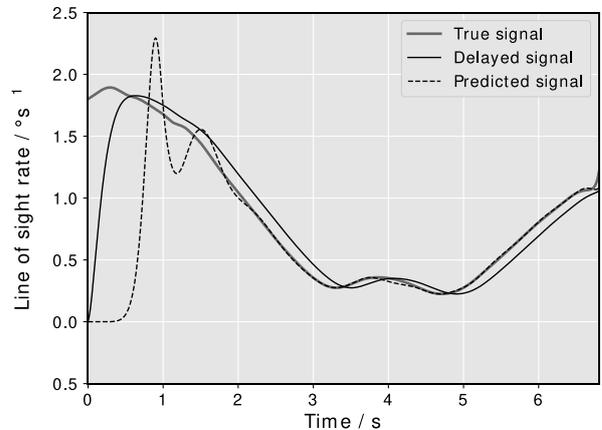

**Figure 4: Line of sight rate for the delayed and predicted seeker signal compared to the true signal**

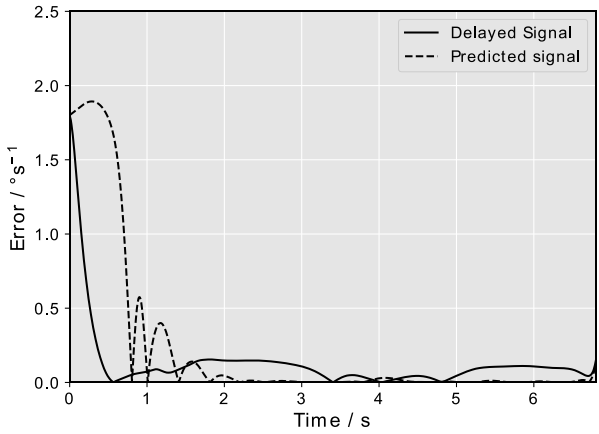

**Figure 5: Error of the delayed and predicted signal compared to the true signal**

Figure 4, demonstrating the tracking of the observer compared the delayed signal highlights the impressive performance of this algorithm. It is apparent that the initial peaking phenomenon, even with the inclusion of a perturbation function, introduces a large error at the start of the guidance period. In this engagement it is likely that this did not affect the miss distance of the simulation because stability was maintained and the observer had converged fully within 2 seconds, significantly before the missile intercepted the target.

Following this period convergence Figure 5 shows the error of the predicted signal remained negligible, particularly when compared to the delayed signal. The result of this accuracy was the large reduction in miss distance that was observed.

As previously stated in the final milliseconds of the engagement the observer deviates from the true signal as it begins to rapidly change. This is particularly apparent from the error plot of the LOS rate, as it is possible to see the increase in the magnitude of the error.

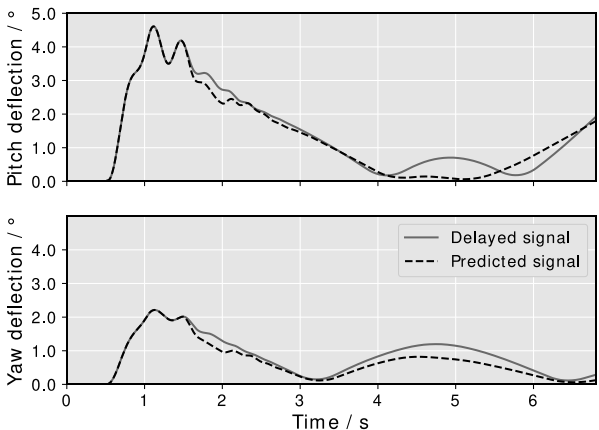

**Figure 6: Control surface deflection of missile model**

Another benefit of introducing the observer was a reduction in the control surface deflection and thus a reduction in the commanded acceleration from the guidance algorithm. The control surface deflections can be seen in Figure 4. After the observer was initialised and began providing the LOS rate signal to the guidance system, the control surface deflection was lower throughout the engagement if the predicted signal was used. Less deflection was observed when the was guidance algorithm correcting the LOS rate as the rocket motor burnt out. This was due to guidance algorithm being able to react earlier to the loss of thrust with the prediction of the observer. A reduction in commanded acceleration would reduce the requirement of the airframe and provide more headroom for the missile in more dynamic engagements.

### 3.2  Randomly Manoeuvring Target

The results of the Monte-Carlo simulation are expressed in Figure 1. The simulation follows a similar exponential growth in miss distance as higher navigation ratios presented by J. Holloway and M. Krstic in their paper.[7] This increases confidence that the simulation result is valid.

The simulation completed successfully for all the scenarios that were executed during the Monte-Carlo simulation. This indicates that the observer maintains a degree of robustness against changing target behaviour and seeker delays. It is important to ensure that this is maintained on real missile platforms and against any countermeasures the missile is expected to encounter but this is an extremely reassuring result in the context of assessing the possible viability of the observer within these systems.

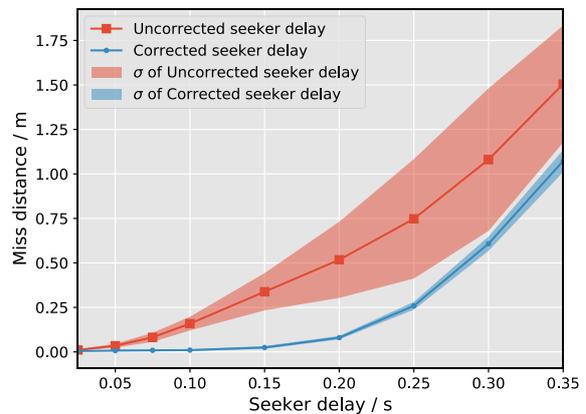

**Figure 7: Monte-Carlo simulation for a randomly manoeuvring target comparing the use of the observer against the use of a delayed sign, includes standard deviation of the miss distances achieved.**

Overall, the randomly manoeuvring target increased the miss distance of the simulation compared to the non-manoeuvring target. This behaviour is expected due to the randomly manoeuvring target being a target model that was designed to be more difficult to intercept to offer a realistic indication of the algorithm's performance against an aircraft.

The results show that there is a significant reduction in the miss distance across the range of delays explored during the simulation. The degree of this improvement is reduced as the delay is reduced because there is less delay to eliminate.

Up to 0.2 seconds delay the miss distance remain flat, which suggest that the observer can eliminate almost all the error caused by seeker delay up to this point. After 0.2 seconds it clearly becomes less effective, but still offers a significant improvement over the incorrected signal.

An unexpected benefit the observer offers is the increase in consistency of the miss distance. The reduction in the standard deviation indicates that the observer will produce a more reliable and accurate result compared to an uncorrected seeker.

## 4. CONCLUSION

The observer based on a two-step differentiator for delayed signal provides excellent delay correction for seeker signals. Integrating the observer into the missile guidance system conferred a large reduction in the miss distance of the missile system for both a non-maneuvering target and a randomly maneuvering target. Furthermore, the inclusion of the observer also increased the consistency of the missile systems performance resulting in less anomalous performances.

The results show that the observer accurately predicts the future signal; the predicted signal produced by the observer is much closer to the true output than the original delayed signal produced by the seeker. This translates to a 54% reduction in the simulated miss distance. A reduction in the commanded acceleration that the guidance algorithm required to be effective reduced the demand on the airframe. It was shown using a Monte-Carlo method that the observer was robust across a range of engagement scenarios, as instability was never presented.

The approach adopted is independent from the dynamics of the system it is integrated within. This is considered to be a positive attribute that could support efficient integration within a real missile system generating significant performance gains to systems that adopt it. This would be particularly relevant to systems if they have large seeker delays, for example, those in mechanical seeker gimbals or introduced during the processing of seeker output.

## REFERENCES

[1] R. Yanushevsky, *Modern missile guidance*. Boca Raton: CRC Press, 2008.
[2] B. A. White and A. Tsourdos, "Modern Missile Guidance Design: An Overview," *IFAC Proc. Vol.*, vol. 34, no. 15, pp. 431–436, 2001.
[3] S. A. Murtaugh and H. E. Criel, "Fundamentals of proportional navigation," *IEEE Spectr.*, vol. 3, no. 12, pp. 75–85, 1966.
[4] N. Dohi, Y. Baba, and H. Takano, "Variable Speed Missile Guidance Law Against a Maneuvering Target," *IFAC Proc. Vol.*, vol. 37, no. 6, pp. 599–604, 2004.
[5] N. Dhananjay, K. Lum, and J. Xu, "Proportional Navigation With Delayed Line-of-Sight Rate," *IEEE Trans. Control Syst. Technol.*, vol. 21, no. 1, pp. 247–253, Jan. 2013.
[6] S. Lee and Y. Kim, "Design of nonlinear observer for strap-down missile guidance law via sliding mode differentiator and extended state observer," in *2016 International Conference on Advanced Mechatronic Systems (ICAMechS)*, 2016, pp. 143–147.
[7] J. Holloway and M. Krstic, "Predictor Observers for Proportional Navigation Systems Subjected to Seeker Delay," *IEEE Trans. Control Syst. Technol.*, vol. 24, no. 6, pp. 2002–2015, 2016.
[8] O. Katsuhiko, *Modern Control Engineering*, 5th ed. Pearson, 2010.
[9] X. Wang and H. Lin, "Two-step differentiator for delayed signal."
[10] U. M. Command, *MIL-HDBK-1211*. Redstone Arsenel, AL, 35898-5270: Department of Defence, 1995.
[11] D. Allerton, *Principles of flight simulation*. Chichester, UK: Wiley, 2009.